# Key Distribution in PKC through Quantas

Aditya Goel

The Technological Institute of Textile & Sciences, Bhiwani, Haryana

adityagoel123@gmail.com

## Abstract

Cryptography literally means "The art & science of secret writing & sending a message between two parties in such a way that its contents cannot be understood by someone other than the intended recipient." and Quantum word is related with "Light". Thus, Quantum Cryptography is a way of describing any information in the form of quantum particles. There are no classical cryptographic systems which are perfectly secure. In contrast to Classical cryptography which depends upon Mathematics, Quantum Cryptography utilizes the concepts of Quantum Physics which provides us the security against the cleverest marauders of the present age. In the view of increasing need of Network and Information Security, we do require methods to overcome the Molecular Computing technologies (A future technology) and other techniques of the various codebrakers. Both the parts i.e. Quantum Key distribution and Information transference from Sender to Receiver are much efficient and secure. It is based upon BB84 protocol. It can be of great use for Govt. agencies such as Banks, Insurance, Brokerages firms, financial institutions, e-commerce and most important is the Defense & security of any country. It is a Cryptographic communication system in which the original users can detect unauthorized eavesdropper and in addition it gives a guarantee of no eavesdropping. It proves to be the ultra secure mode of communication b/w two intended parties.

Final Paper

## (1). Introduction to Quantum Cryptography: Quantum Cryptography

provides a completely secure way to solve the problem of Key Distribution and to transmit any message data safely. It utilizes the Rule of Quantum Physics[8] i.e.

"Heisenberg uncertainty Principle", according to which measuring a quantum particle in

general disturbs it and yields incomplete information about its state before the

measurement. Generally, the quantum Particles used is a short burst of light called as the

Photons because their behavior is comparatively well-understood as well as they can be sent through a fiber optic cable rather easily which is the most suitable medium for extremely high-bandwidth communications. Also a photon has a property called polarization and that property can be measured. The polarization of a particle is the direction in which the wave is oscillating or the angle of the vibration is known as the polarization of the photon. [2] [7] The polarization can be measured either rectilinearly (UP/DOWN $\{90^0\}$, LEFT/RIGHT $\{0^0\}$) or diagonally (UPLEFT/RIGHTDOWN $\{135^0\}$ and UPRIGHT/LEFTDOWN $\{45^0\}$). These are two bases of measurement i.e. Rectilinear and Diagonal. Each bit of the data to be transferred will be in the form of a photon of the light. Eavesdropping on a quantum communication channel therefore causes an unavoidable disturbance, alerting the genuine/original users. This key can then be used with any chosen encryption algorithm to encrypt (and decrypt) a message. The algorithm most commonly associated with QKD is the one time pad, as it is secure when used with a secret, random key. Quantum cryptography has limitations like the photon keys can travel only between computers directly connected through fiber-optic lines (as opposed to networked systems), and the photons degenerate over long distances but with the advancements in technology they are assumed to fade dim.

**(2).Working Of Quantum Cryptography[15]:** Many Models have been presented to describe the working of the Quantum Cryptography. All protocols have been applied experimentally. The famous protocols among them are:

➤ BB84 [7][9] - Relies on Weisner's[10] "Conjugate coding" proposal and uses four polarization states. It was developed by *B*ennett and *B*rassard. It was fully implemented in 1989 most prevalent.

➤ E91 protocol [11][12]- Relies on EPR entanglement uses Bell's Inequality and quantum entanglement. It was developed by Artur Ekert.

➤ B92 [13]- Similar to BB84 but only uses two polarization states.

➤ Interference phase drift - This protocol uses the mechanism of interference to ensure security. Eavesdropping would cause any interference to be destroyed, and so can be detected.

**Working of BB84 protocol:** *B*ennett and *B*rassard presented the working procedure of Quantum Cryptography in 19*84* commonly known as BB84 protocol after its inventors and year of publication, but the full implementation of the above was done in 1989 at the Thomson J. Watson Research Center of IBM in Yorktown Heights, New York, USA. It has been demonstrated for a distance of 30 Kms over a fiber optic link and for a distance of 100 mts using air as a media.

According to this Protocol, It uses two channels, one quantum and one simple. It is assumed that three people are involved Sender of the information, Receiver and eavesdropper who wants to listen or to crack the information exchange between the Sender and Receiver. He also can easily see and hear everything that takes place over the 2nd insecure channel. All three have access to both channels but not to each other's computers. Eavesdropper cannot corrupt information on the public channel and he cannot break the laws of physics. However he is assumed to have unlimited computing power. In order to send an encrypted message Sender first communicates her intent with Receiver. He then sends the Key which will be the mode for encrypting the message data. Key is sent in the form of stream of individual photons provided each bit can have value either 1 or 0, through the quantum channel. However, in addition to their linear travel, all of these photons are oscillating (vibrating) in a certain manner. As explained earlier, Oscillations here are assumed to be of four types such as at $0^0$, $45^0$, $90^0$ or $135^0$. A polarizer is simply a filter that permits photons to pass through it with any angle out of four explained oscillations. Sender has a polarizer that can transmit the photons in any one of the four states mentioned. It totally depend upon the orientation of the polarizer that either it is a Rectilinear or Diagonal polarizer. Also Sender swaps his polarization scheme between rectilinear and diagonal filters for the transmission of each single photon bit in a random manner so that a continuous altering stream of bits are created. In doing so, the transmission can have one of two polarizations represent a single bit, either 1 or 0, in either scheme she uses. He records this random sequence. In the Diagram shown above, every photon (acting as a bit) may be assigned either as 0 or 1. (Refer to the Diagram below.)[7]

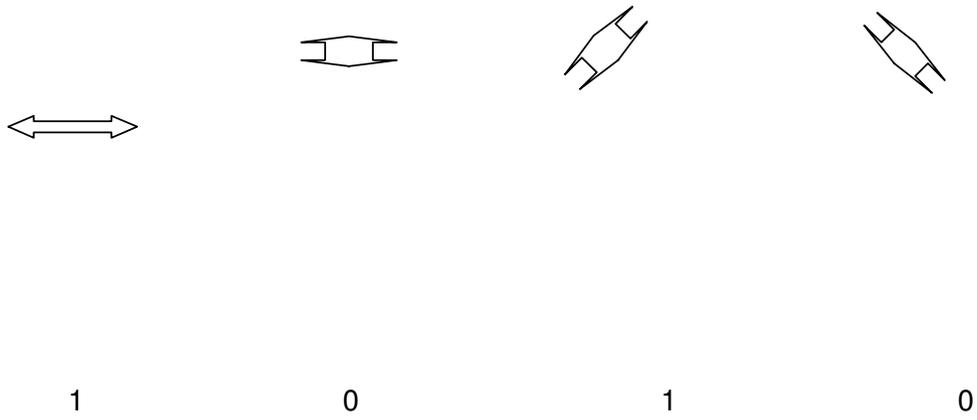

|   |   |   |   |
|---|---|---|---|
| 1 | 0 | 1 | 0 |

When receiving the photon key, Receiver must choose to measure each photon bit using either his rectilinear or diagonal polarizer: sometimes he will choose the correct polarizer and at other times he will choose the wrong one. Like Sender, he selects each polarizer in a random manner. Now we can make a guess as to what happens with those photons for which wrong orientation of the detector (Wrong polarizer) was used. Suppose Receiver uses a rectilinear polarizer to measure UPLEFT/ RIGHTDOWN and UPRIGHT/ LEFTDOWN (diagonal) photons. If he does this, then the photons will pass through in a changed state - and others incorrectly. At this point, Receiver and Sender must establish a $2^{nd}$ channel of communication that may be insecure for conforming to which polarizer he used to send each photon bit - but not how she polarized each photon. So, he could say that photon number 7854 was sent using the rectilinear scheme, but he will not say whether he sent an UP/DOWN or LEFT/RIGHT. Receiver then confirms if he used the correct polarizer to receive each particular photon bit. They then discard all the photon measurements for which Receiver used the wrong polarizer to measure. Of those cases where Receiver choose the correct detector are translated into bits (1's and 0's) and he can then determine a binary key.

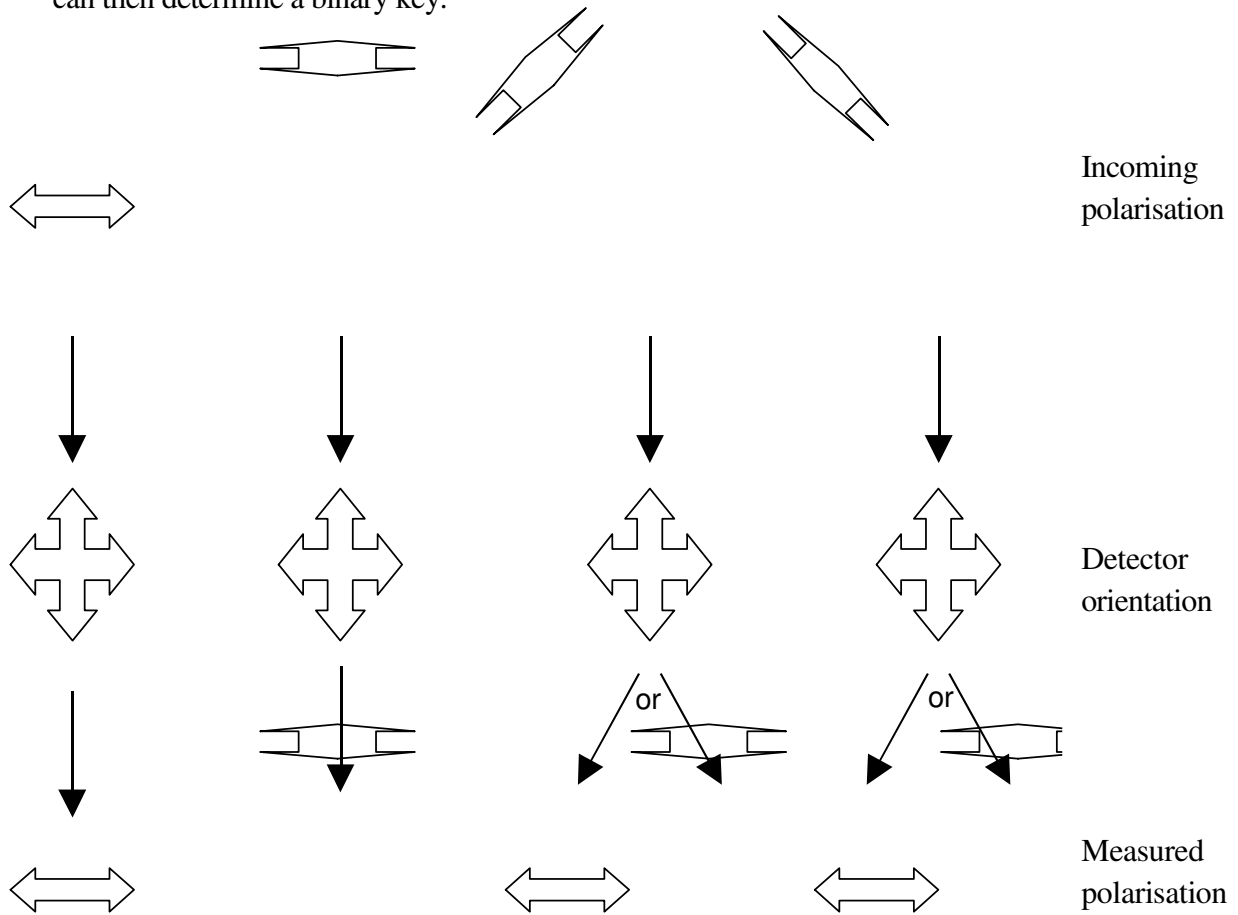

Once the information has been collected Sender and Receiver must then use a sequence of error corrections on the data in order to eliminate the effects of both bad channels and eavesdropping. Once the binary key has been formed, it will form the basis for Vernam Cipher, the only cryptosystem that, if properly implemented, is proven to be completely random and secure. If any errors were found in the raw quantum key the key could be assumed to have been compromised and could be abandoned [BB84]. However further work by those two in collaboration with Jean-Marc Robert (a student of Brassard), has led to the development of a system known as privacy amplification [BBR88][16]. This assumes that Sender and Receiver share a secret key of length $p$ and that eavesdropper knows some subset of these bits of size $l$ where $l < p$. Sender and Receiver don't know which of these bits have been compromised. Sender and Receiver can estimate from various methods what amount of the key eavesdropper may have gained from intercept/resend and from beam splitting. From there they can use privacy amplification to give them a key of which eavesdropper can know nothing.

## Attacks On BB84 Protocol[14]:

Now, suppose we have an eavesdropper, who attempts to listen in, has the same polarizers that Receiver have. He has at her disposal two main strategies for trying to steal information from the quantum channel.

> Intercepting / resending. Eavesdropper would have to choose whether he wanted to record rectilinearly or diagonally without knowing which way Sender had aligned her transmitters. It is possible that eavesdropper has correctly guessed the alignment of the photo detectors and has therefore got accurate results. He can then send brand new photons at this polarization on to receiver. If however he has chosen the wrong alignment then he would have been detected. Eavesdropper will then transmit this bit in the polarization in which he detected it. It can be shown experimentally that 25% of the bits eavesdropper intercepts will yield the wrong results if transmitted on to Receiver. This is usually referred as **intrusion detection**. To demonstrate the fact that Eavesdropper is on the "photon highway"

Let us take an example: Let's say that Sender transmits photon number 349 as an UPRIGHT/LEFTDOWN to Receiver, but for that one, eavesdropper uses the rectilinear polarizer, which can only measure UP/DOWN or LEFT/RIGHT photons accurately. What eavesdropper will do is transform that photon into either UP/DOWN or LEFT/RIGHT, as that is the only way the photon can pass. If Receiver uses his rectilinear polarizer, then it will not matter what he measures as the polarizer check When Sender and Receiver go through the public channel communication they will discard that photon from the final key. But if eavesdropper uses the diagonal polarizer, a problem arises & he may measure it correctly as UPRIGHT/LEFTDOWN, but he stands an equal chance, according to the Heisenberg Uncertainty Principle, of measuring it and changing it to some other state which would also be detected.

➢ Beam splitting would require putting some form of half-silvered mirror in front of the beam in order to split it into two. The first half of the beam would travel onto Receiver and Eavesdropper could analyze the second part. This method of attack would, in a perfect system, *be impossible* because the pulse used would be exactly one photon and so it would be impossible to split. Beam splitting also dramatically increases the error rate at Receiver's end of the channel. It is assumed that Eavesdropper can store this pulse of light for as long as needed until Sender declares her sending polarizations and Eavesdropper can read them in what best way he thinks. Also it would result into the no stream of bits at the original receiver's destination.

**The E91 protocol:** The Ekert scheme uses entangled pairs of photons. These can be created by Sender, by Receiver, or by some source separate from both of them, including eavesdropper. The photons are distributed so that Sender and Receiver each end up with one photon from each pair. The scheme relies on two properties of entanglement.

➢ First, the entangled states are perfectly correlated in the sense that if Sender and Receiver both measure whether their particles have vertical or horizontal polarizations, they will always get the same answer with 100% probability.

➢ Second, any attempt at eavesdropping by Eavesdropper will destroy these correlations in a way that Sender and Receiver can detect.

Researchers from Northwestern University and Massachusetts-based BBN Technologies demonstrated fully-functional quantum cryptographic data network which leverages the quantum entanglement properties of photons for both data transfer as well as key distribution (E91 Protocol). This current breakthrough combined Northwestern's data encryption method (known as AlphaEta) with BBN's key encryption scheme to enable a completely secure fiber optic link between BBN's headquarters and Harvard University, a distance of nine kilometers. As you might imagine, the entire project was funded by a $5.4 million grant from DARPA (Defense Advanced Research projects Agency Of US.), an agency which has a vested interest in transmitting data that not even a theoretical quantum computer could crack.

## (5). Organizations Involved for the Implementation of the Quantum Cryptography:  In 2004, the world's first bank transfer using quantum cryptography was carried in Vienna, Austria. An important cheque, which needed absolute security, was transmitted from the Mayor of the city to an Austrian bank.

The DARPA Quantum Network has been running since 2004 in Massachusetts, USA. It is being developed by BBN Technologies, Harvard University and Boston University.

There are currently four companies [17]offering commercial quantum cryptography systems; id Quantique (Geneva), MagiQ Technologies (New York), SmartQuantum (France) and Quintessence Labs (Australia). Several other companies also have active research programmes, including Toshiba, HP, IBM, Mitsubishi, etc.

Researchers from Northwestern University and Massachusetts-based BBN Technologies demonstrated fully-functional quantum cryptographic data network which leverages the quantum entanglement properties of photons for both data transfer as well as key distribution (E91 Protocol). This current breakthrough combined Northwestern's data encryption method (known as AlphaEta) with BBN's key encryption scheme to enable a completely secure fiber optic link between BBN's headquarters and Harvard University, a distance of nine kilometers

As of March 2007 distance over which quantum key distribution has been demonstrated using optic fibre is 148.7 km, achieved by Los Alamos/NIST using the BB84 protocol. Significantly, this distance is very long and the distance record for free space QKD is 144 km between two of the Canary Islands, achieved by a European collaboration using entangled photons (the Ekert scheme). Quantum encryption system provided by the Swiss company Id Quantique was used in the Swiss canton (state) of Geneva to transmit ballot results to the capitol in the national election occurring on Oct. 21, 2007. The world's first computer network protected by quantum cryptography was implemented in October 2008, at a scientific conference in Vienna. The network used 200 km of standard fiber optic cable to interconnect six locations across Vienna and the town of St. Poelten located 69 km to the west. The eavesdropper was witnessed by Gilles Brassard and Anton Zeilinger.

## Conclusion

From the view point of the Information and Network security, Quantum Cryptography is the most advanced and most comprehensive way of safely transferring the data between intended parties. It is based upon the Heisenberg's Uncertainty Principle which sits at the heart of Quantum Mechanics. It not only detects the eavesdropper detecting the secret information but also transmits him the wrong information. Sir Stephner in 1970 has coined this term first by writing a paper on the Conjugate Coding. Then Bennett and Brassard worked upon it and came with the full working model in 1989 by the help of various other personalities. It is very expensive technology but along with the time, we can expect this disadvantage to be disappeared. It still has to be implemented in a better way in various developing well as developed countries. It currently uses Fiber optic link to be implemented. There is still a very large scope in it's improvement to be implemented through the air as an medium. It represents the next line of IT security.